\documentclass[twocolumn]{revtex4}
\pdfoutput=1
\usepackage{amsfonts}
\usepackage{amsmath}
\usepackage{graphicx}
\usepackage{dcolumn}
\usepackage{bm}
\usepackage{amssymb}

\begin{document}

\title{Measuring Temperature Gradients over Nanometer Length Scales}


\author{Eric A. Hoffmann}
\email{ehoffma1@uoregon.edu}
\affiliation{Physics Department and Materials Science Institute, University of Oregon, Eugene, Oregon 97403-1274}
\author{Henrik A. Nilsson}
\affiliation{Solid State Physics/The Nanometer Structure Consortium, Lund University, Box 118, S-221 00, Lund, Sweden}
\author{Jason E. Matthews}
\author{Natthapon Nakpathomkun}
\author{Ann I. Persson}
\affiliation{Physics Department and Materials Science Institute, University of Oregon, Eugene, Oregon 97403-1274}
\author{Lars Samuelson}
\affiliation{Solid State Physics/The Nanometer Structure Consortium, Lund University, Box 118, S-221 00, Lund, Sweden}
\author{Heiner Linke}
\email{linke@uoregon.edu}
\affiliation{Physics Department and Materials Science Institute, University of Oregon, Eugene, Oregon 97403-1274}

\begin{abstract}
When a quantum dot is subjected to a thermal gradient, the temperature of electrons entering the dot can be determined from the dot's thermocurrent if the conductance spectrum and background temperature are known. We demonstrate this technique by measuring the temperature difference across a 15 nm quantum dot embedded in a nanowire. This technique can be used when the dot's energy states are separated by many $kT$ and will enable future quantitative investigations of electron-phonon interaction, nonlinear thermoelectric effects, and the efficiency of thermoelectric energy conversion in quantum dots.
\end{abstract}

\maketitle

Quantum and finite-size effects modify the thermal properties of electronic mesoscopic devices. Examples of novel thermal and thermoelectric phenomena demonstrated by mesoscopic devices include the quantization of the electronic thermal conductance\cite{Molenkamp92}, energy-modulated thermovoltage\cite{Molenkamp94} in quantum point contacts (QPCs), and nonlinear phenomena, such as thermal diode behavior in quantum dots\cite{Molenkamp08}. Applications of mesoscopic thermal effects include the use of low-dimensional systems with sharp features in the electronic density of states for efficient thermoelectric energy conversion \cite{Mahan96}. For example, quantum dots\cite{Humphrey02, O'Dwyer06} and nanowires\cite{Hicks03, Heath08} are being explored for use as efficient thermoelectric devices.

A challenge when performing a low-temperature, quantitative, mesoscopic thermal experiment is the measurement of the temperature difference across the small device under conditions where electronic and lattice temperatures can be very different. In the field of mesoscopic thermoelectrics, such thermometry is often accomplished using the thermovoltage of a QPC to measure temperature differences within a two-dimensional electron gas. \cite{Molenkamp92,Molenkamp94,Molenkamp07,Nicholls08}. Thermal experiments using micrometer-length nanotubes and nanowires often exploit the temperature dependence of local resistors to determine the lattice temperature difference across the device\cite{Hone98,Kim03,Shi03,Llaguno04,Seol07,Heath08}.

Recently, we proposed a novel thermometry technique\cite{Hoffmann07}, which uses a two-terminal quantum dot to measure separately the electron gas temperature on the source and drain sides of a quantum dot that has been placed in a temperature gradient. Crucially, this technique measures the temperature of precisely those electrons which enter the dot, rather than the temperature of separate electrons (or phonons) in the vicinity. This can be a significant advantage, for example, in the study of thermal phenomena in quantum dots that depend on electronic temperature, such as inelastic processes or thermoelectric phenomena. In such experiments, the use of quantum-dot thermometry significantly simplifies device layout and fabrication, because the same quantum dot is used for both experiment and thermometry. An example application of this local thermometry technique is to embed a quantum dot into a nanowire in order to answer fundamental questions about the nature of electron-phonon (e-ph) interaction and its role in heat flow in nanowire-based devices.

\begin{figure}[t!]
\begin{center}
\leavevmode\includegraphics{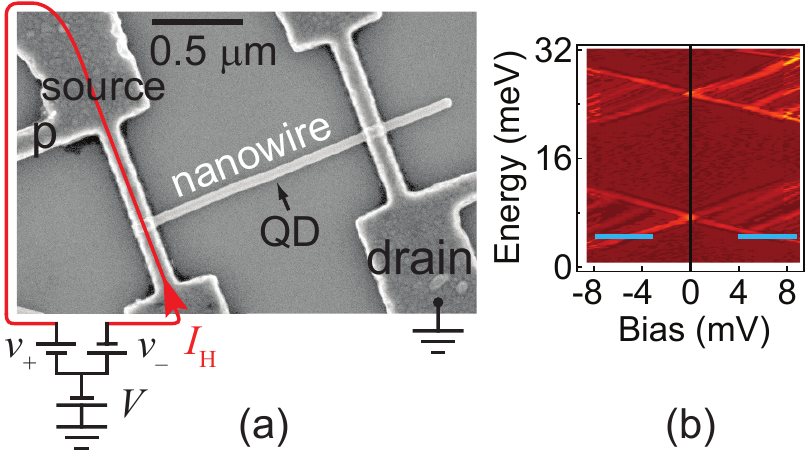}
\end{center}
\caption{a) An SEM image of the InAs nanowire with source and drain contacts. The indicated InP/InAs/InP quantum dot (QD) embedded in the nanowire is not resolved by the SEM. The voltage, $V$, biases the nanowire electrically, and the heating current, $I_{\text{H}}$, biases the nanowire thermally. The voltage probe (p) assists in tuning the heating voltages, $v\pm$ (see text for details). b) Differential conductance as a function of bias voltage and gate energy with brighter color indicating larger differential conductance. The dot is in the few-electron Coulomb-blockade regime and, in the center Coulomb diamond, is filled with an odd number of electrons. The average charging energy of the dot is 12.9 meV. The (blue) horizontal lines indicate regions where the assumptions leading to Eq.~(\ref{Signal Ratio}) are satisfied and at which gate and bias values the thermometry data was taken (see Figs. \ref{stack} and \ref{trend}).}
\label{device}
\end{figure}

\begin{figure}[t!]
\begin{center}
\leavevmode\includegraphics{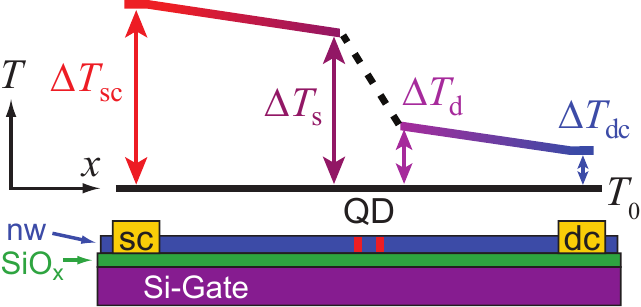}
\end{center}
\caption{
A schematic (not to scale) of the temperature profile along a nanowire (nw) with an embedded quantum dot (QD), a heated metallic source contact (sc), and an unheated metallic drain contact (dc). In the source and drain contacts, the electron gas temperature rises by $\Delta T_{\text{sc}}$ and $\Delta T_{\text{dc}}$ above the cryostat temperature, $T_{\text{0}}$, respectively. In the nanowire, the electron gas temperature rises by $\Delta T_{\text{s,d}}$ near the source and drain sides of the quantum dot, respectively ($\Delta T_{\text{sc}} \ge \Delta T_{\text{s}} \ge \Delta T_{\text{d}} \ge \Delta T_{\text{dc}}$).}
\label{temperature}
\end{figure}

Here we present the experimental demonstration of quantum-dot thermometry. Specifically, we use a quantum dot defined by a double barrier within a nanowire that is placed on an electrically insulating substrate and to which metal Ohmic contacts are attached (see Fig.~\ref{device}a). During thermometry measurements, a temperature difference and bias voltage, $V$, are applied along the nanowire and across the quantum dot. A heating current, $I_{\text{H}}$, heats the electron gas in the metallic source contact (sc) while the electron gas in the metallic drain contact (dc) is expected to stay at or near the background cryostat temperature, $T_{\text{0}}$, creating a temperature difference over the length of the nanowire (Fig.~\ref{temperature}). In the following, we discuss how we measure the temperature rises, $\Delta T_{\text{s,d}}$, above $T_{\text{0}}$ experienced by the electron gas inside the nanowire near the source and drain sides of the quantum dot, respectively (see Fig.~\ref{temperature}).

Our method makes use of the fact that, when the nanowire drain is electrically grounded, the applied temperature difference, $\Delta T_{\text{s}}-\Delta T_{\text{d}}$, causes a net electron charge current, $I_{\text{th}}$, known as the \textit{thermocurrent}. Under appropriate gate and bias conditions (see Fig.~\ref{device}b), $I_{\text{th}}$ depends only on the temperature of the electrons entering the dot, and on the dot's energy-dependent transmission function, $\tau \left(\varepsilon\right)$. Transmission information can be obtained from conductance spectroscopy; specifically, the second differential conductance, $\partial ^{2}I/\partial V^{2}$, provides the necessary information about the dot's transport properties, and a mathematical comparison of $I_{\text{th}}$ to $\partial ^{2}I/\partial V^{2}$ determines $\Delta T_{\text{s,d}}$. The original theoretical proposal\cite{Hoffmann07} for this method considered the regime in which the width of the dot's transmission resonances, $\Gamma$, is larger than the thermal energy $kT$ and where, for a given $\Gamma$, a numerically determined integration factor is needed to determine $\Delta T_{\text{s,d}}$.

\begin{figure}[t]
\begin{center}
\leavevmode\includegraphics{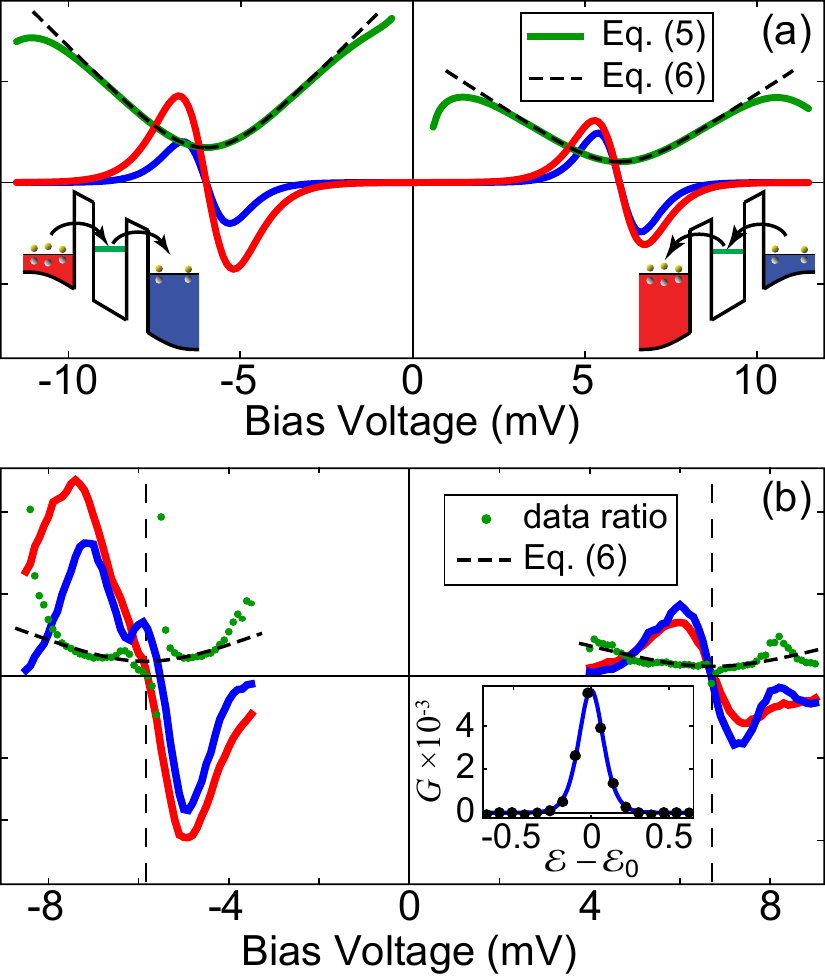}
\end{center}
\caption{a) Modeled thermocurrent, $I_{\text{th}}$, (red) and second differential conductance, $\partial^{2}I/\partial V^{2}$, (blue) as a function of bias voltage calculated using Eq.~(\protect\ref{Landauer}) and a Lorentzian $\tau\left(\varepsilon\right)$ with $\Gamma = 10$ $\mu \text{eV} \ll kT=190$ $\mu \text{eV} = 2.2$ K. Near a resonant energy (see insets), both $I_{\text{th}}$ and $\partial^{2}I/\partial V^{2}$ have a zero-crossing, and their ratio (green) behaves according to Eq.~(\ref{Signal Ratio}) (dashed). \textit{Insets}: When biased negatively, heated electrons flow through the dot, while cold electrons flow when positively biased. b) Experimental data of $I_{\text{th}}$ (red), $\partial ^{2}I/\partial V^{2}$ (blue), their ratio (green), and theory Eq.~\ref{Signal Ratio} (dashed) as a function of bias voltage. For these data, $I_{\text{H}}=150$ $\mu$A, $\Delta T_{\text{s}}=230$ mK, $\Delta T_{\text{d}}=160$ mK, and $T_{\text{0}}=2.2$ K. \textit{Inset}: Differential conductance, $G$, measured at $T_0 = 550$ mK in units of $2e^{2}/h \times 10^{-3}$ as the resonant energy of the dot is swept via the back-gate. The (black) dots are experimental data, and the solid (blue) line is a theoretical fit (see text for details).}
\label{data}
\end{figure}

Here we report an experiment in the regime where $\Gamma \ll kT$. In this limit, $\Delta T_{\text{s,d}}$ can be determined directly from the ratio $I_{\text{th}}/\left( \partial ^{2}I/\partial V^{2}\right)$, without numerical calibration. To demonstrate this, we begin with the Landauer equation for the two-terminal current through a one-dimensional constriction\cite{Datta} 
\begin{equation}
I=\frac{2e}{h}\int_{-\infty }^{\infty }\left[ f_{\text{s}}\left( \varepsilon,V
\right) -f_{\text{d}}\left( \varepsilon,V \right) \right] \tau \left(
\varepsilon\right) d\varepsilon,  \label{Landauer}
\end{equation}
where $f_{\text{s,d}}^{-1}=e^{\xi _{\text{s,d}}}+1$ are the Fermi-Dirac
distributions in the source and drain, respectively. We assume that the bias voltage,
$V$, drops symmetrically by $V/2$ across each barrier so that
$\xi _{\text{s,d}}=\left( \varepsilon -\mu \pm eV/2\right) /kT_{\text{s,d}}$, where $\mu$ is controlled by the gate voltage. 
In the following, we assume that transmission resonances are separated by many $kT$ and that $\Gamma \ll kT$. We consider a single resonance, which is centered at energy $\varepsilon_{\text{0}}$ and is so sharp that $\tau\left(\varepsilon\right)$ limits to a Dirac delta function, $\tau \left( \varepsilon \right)=A\delta \left( \varepsilon -\varepsilon _{\text{0}}\right)$, where $A$ is an unknown constant. Inserting this expression for $\tau \left( \varepsilon \right)$ into Eq.~(\ref{Landauer}) and integrating gives
\begin{equation}
I=A\frac{2e}{h}\left[ f_{\text{s}}\left( \varepsilon _{\text{0}},V\right) -f_{
\text{d}}\left( \varepsilon_{\text{0}},V\right) \right] .
\label{Dirac Landauer}
\end{equation}
In general, current through the nanowire depends on both $T_{\text{s}}$ and $T_{\text{d}}$. However, using the gate and bias voltages, one can align either the source or drain electrochemical potential with the resonance at $\varepsilon_{\text{0}}$, while keeping the opposite electrochemical potential several $kT$ from any resonance (see Fig.~\ref{device}b and insets of Fig.~\ref{data}a). In such a configuration, only one Fermi-Dirac distribution in Eq.~(\ref{Dirac Landauer}) is finite near $\varepsilon_{\text{0}}$, and current through the dot depends on only one temperature:~$T_{\text{s}}$ or $T_{\text{d}}$. Therefore, under these single-temperature bias conditions, the thermocurrent, $I_{\text{th}}$, can be written
\begin{equation*}
\left. I_{\text{th}}\right\vert _{\text{s,d}}=\Delta T_{\text{
s,d}}\frac{\partial I}{\partial T_{\text{s,d}}},
\end{equation*}
where the subscripts denote which respective electrochemical potential, source or drain, is aligned with the resonance. After differentiating Eq.~(\ref{Dirac Landauer}) with respect to $T_{\text{s,d}}$, $I_{\text{th}}$ becomes 
\begin{equation}
\left.I_{\text{th}}\right\vert_{\text{s,d}}=\Delta T_{\text{s,d}}A\frac{2e}{h}\left[\mp\frac{\partial f_{\text{s,d}}}{\partial \xi_{\text{s,d}}}\frac{\xi_{\text{s,d}}}{T_{\text{s,d}}}\right].\label{diff. thermo.}
\end{equation}
This alone is not sufficient to determine $\Delta T_{\text{s,d}}$ because $A$ is unknown.  However, the second differential conductance, $\partial^{2}I/\partial V^{2}$, provides additional information and resolves this issue. Under the same single-temperature bias conditions, and again using Eq.~(\ref{Dirac Landauer}), the second differential conductance is
\begin{equation}
\left.\frac{\partial^{2}I}{\partial V^{2}}\right\vert_{\text{s,d}}=\frac{e^{2}}{4k^{2}}\frac{1}{T_{\text{s,d}}}A\frac{2e}{h}\left[\pm\frac{\partial f_{\text{s,d}}}{\partial\xi_{\text{s,d}}}\frac{2f_{\text{s,d}}-1}{T_{\text{s,d}}}\right].\label{2nd diff. cond.}
\end{equation}
The bracketed parts of Eqs.~(\ref{diff. thermo.}) and (\ref{2nd diff. cond.}) have terms in common. This is the primary reason for comparing $I_{\text{th}}$ to $\partial^{2}I/\partial V^{2}$ rather than the (conventional) differential conductance, $\partial I/\partial V$. In particular, we consider the ratio
\begin{equation}
R\equiv I_{\text{th}}/{\frac{\partial ^{2}I}{\partial V^{2}}}.
\label{Gen Ratio}
\end{equation}
To approximate $R$ under the single-temperature bias conditions, we define $R_{\text{s,d}}$ as the ratio of Eq.~(\ref{diff. thermo.}) over Eq.~(\ref{2nd diff. cond.})
\begin{equation*}
R\approx R_{\text{s,d}}\equiv \left.I_{\text{th}}/{\frac{\partial ^{2}I}{\partial V^{2}}}\right\vert_{\text{s,d}}= -\Delta T_{\text{s,d}}\frac{{4k}^{2}}{e^{2}}T_{\text{s,d}}\frac{\xi _{\text{s,d}}}{2f_{\text{s,d}}-1},
\end{equation*}
and $A$ conveniently drops out. By expressing this ratio in bias voltage and defining the voltage values $V_{\text{s,d}}^{\text{0}} \equiv \mp2\left(\varepsilon_{\text{0}}-\mu\right)/e$, $R_{\text{s,d}}$ can be written
\begin{equation}
R_{\text{s,d}}=\Delta T_{\text{s,d}}\frac{2k}{e}\left(V-V_{\text{s,d}}^{\text{0}}\right) \coth \left(\frac{e}{4k}\frac{V-V_{\text{s,d}}^{\text{0}}}{\Delta T_{\text{s,d}}+T_{\text{0}}}\right).\label{Signal Ratio}
\end{equation}
This equation is the main theoretical result of this Letter. Note that it is only valid when either the source or drain electrochemical potential is near a resonance of the quantum dot, but not both, and when $\Gamma \ll kT$. In the remainder of the paper we will discuss our experimental device and setup, present experimental data, explain how $\Delta T_{\text{s,d}}$ are extracted from the data, and conclude with comparison to numerical modeling.

The experimental device (see Fig.~\ref{device}a) is an InAs nanowire about 55 nm in diameter and 1.2 $\mu$m long grown using chemical beam epitaxy. Two 5 nm InP barriers embedded during growth define an InAs quantum dot about 15 nm in length. After growth, the nanowire is deposited onto an n-doped Si wafer capped with a 100 nm electrically insulating SiO$_{\text{x}}$ layer. The n-doped wafer serves as a back-gate (see Fig.~\ref{temperature}), and Ni/Au electrical contacts are defined at both ends of the nanowire using electron beam lithography. For further growth and fabrication details, see Ref.~\cite{Thelander04}. Electron-electron Coulomb interaction splits degenerate quantum energy levels into single-electron energy states (the so-called Coulomb blockade). The classical charging energy is on average 12.9 meV (see Fig.~\ref{device}b), and the quantum energy spacing is as large as 40 meV. In all experiments, $T_{\text{0}} < 5$ K $\cong 0.4 \text{ meV}$ guaranteeing that the energy levels are spaced in energy by much more than the thermal energy. Near energy resonances (quantum energy levels split by Coulomb blockade), the transmission function of the double-barrier quantum dot is Lorentzian\cite{Ferry&Goodnick}. The InP barriers provide long electron lifetimes and make each Lorentzian very sharp. In fact, theoretical fits of differential conductance peaks that assume an infinitely sharp Lorentzian (see Eq.~(1) in Ref.~\cite{Foxman93}) fit our data very well. The inset of Fig.~\ref{data}b is such a fit and uses $T_{\text{0}}$ (measured with a calibrated Cernox sensor heat sunk in the same way as the device) and the capacitance ratio $\alpha=C_{\text{g}}/C_{\Sigma}$ (determined from Coulomb Blockade diamonds). The center of the transmission resonance is the only fitting parameter. The fits indicate that temperature smearing is the dominant peak-broadening mechanism over our experimental background temperature range. Therefore, in our analysis below, we assume $\Gamma \ll kT_{\text{0}}$ and the validity of Eq.~(\ref{Signal Ratio}).

In mesoscopic experiments, it is standard practice to use an ac heating current $I_{\text{H}}$, and take advantage of lock-in amplification techniques in order to improve the signal-to-noise ratio. We use a lock-in amplifier to measure the frequency-doubled electrical current which flows in response to $I_{\text{H}}$. This current is the \textit{differential} thermocurrent, $I_{\text{th}}$, because it is the \textit{change} in electrical current which flows as a result of the \textit{change} in the thermal gradient. In the experiment, the ac heating current is created by applying two heating voltages, $v_{\pm}$, which are out of phase with each other by one-half cycle. The relative amplitudes of the two heating voltages are tuned so that their sum is zero at the nanowire's source contact. This voltage balancing prevents $v_{\pm}$ from biasing the nanowire. In practice, their sum cannot be tuned to exactly zero. A concern is that second-order nonlinear processes create a frequency-doubled current, which appears as a phantom thermocurrent. However, quantitatively this nonlinear signal is two orders of magnitude smaller than the thermocurrent and behaves qualitatively different than the thermocurrent traces. Therefore, the effect of the small residual heating voltage at the nanowire is negligible. For the data taken at 2.94 K, $I_{\text{H}}$ has an rms amplitude of up to $460 \mu$A and a frequency of 62.5 Hz. Concurrent with measurements of $I_{\text{th}}$, we apply an ac bias with an amplitude of 67 $\mu$V and a frequency of 40 Hz and measure $\partial I/\partial V$ using a lock-in amplifier.
$\partial^{2}I/\partial V^{2}$ is then calculated by taking a numerical derivative.
\begin{figure}[t]
\begin{center}
\leavevmode\includegraphics{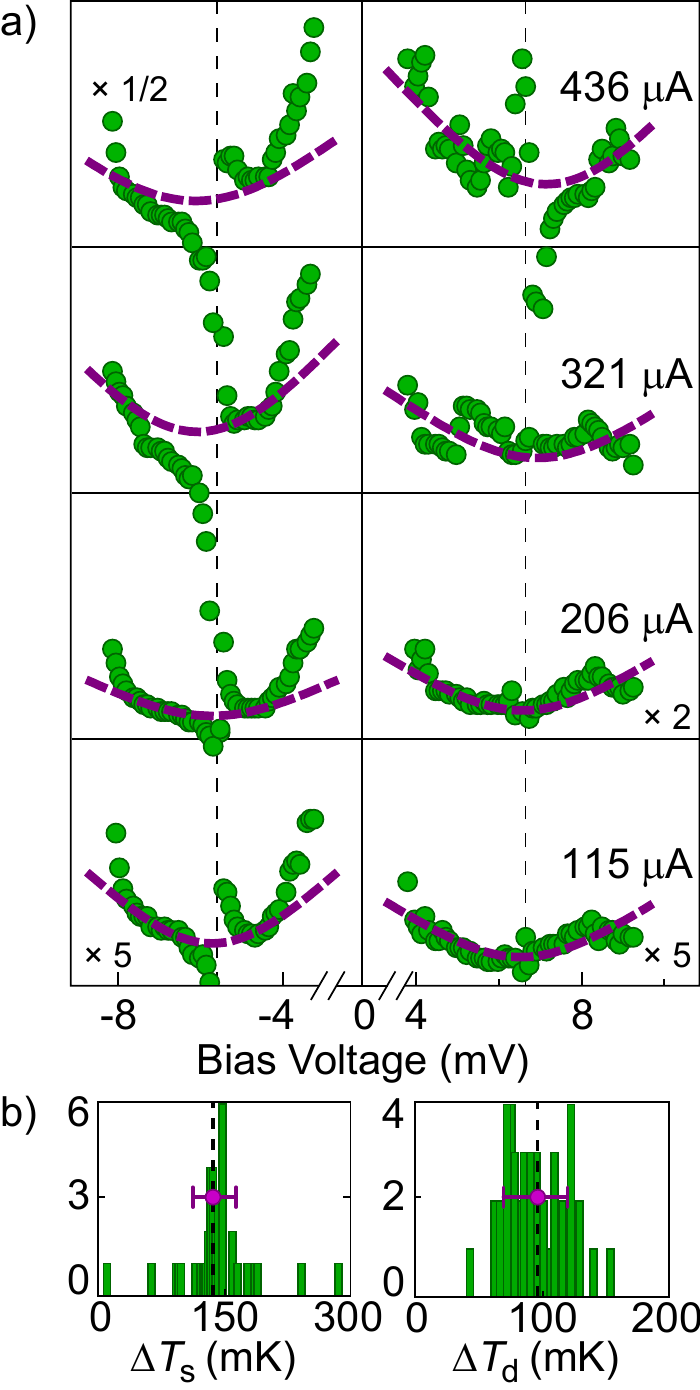}
\end{center}
\caption{a) The (green) data points are the ratio $R=I_{\text{th}}/\left(\partial^{2}I/\partial V^{2}\right)$ measured at four different heating currents, $I_{\text{H}}$, as indicated and at $T_{\text{0}}=2.94$ K. The left data sets determine $\Delta T_{\text{s}}$, and the right sets $\Delta T_{\text{d}}$. The (violet) dashed lines are Eq.~(\ref{Signal Ratio}) based on the mean of the histograms in (b). The data sets and theory have been multiplied and/or offset for clarity. b) Histograms of the values of $\Delta T_{\text{s,d}}$, obtained by solving Eq.~(\ref{Signal Ratio}), at $I_{\text{H}}=115 \mu$A shown in (a). The dot indicates the mean of the data, and the error bars encompass 67\% of the histogram data.}
\label{stack}
\end{figure}

\begin{figure}[t!]
\begin{center}
\leavevmode\includegraphics{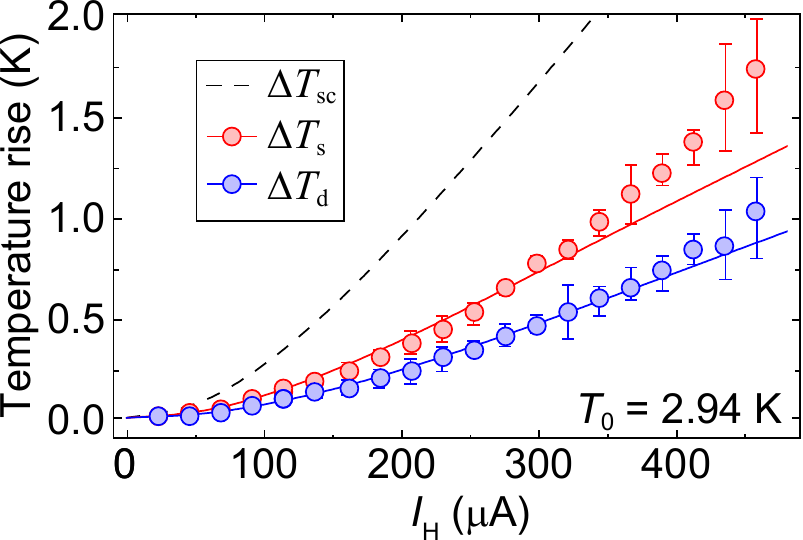}
\end{center}
\caption{A plot of the temperature rises, $\Delta T_{\text{sc}}$, $\Delta T_{\text{s}}$, and $\Delta T_{\text{d}}$ (see Fig.~\ref{temperature}) measured at $T_{\text{0}}=2.94$ K as a function of the heating current, $I_{\text{H}}$. The circular data points and error bars are obtained from histograms (see Fig.~\ref{stack}). The dashed line is a numerical calculation of $\Delta T_{\text{sc}}$. (The change in drain temperature is negligible.) The solid lines are numerical calculations of $\Delta T_{\text{s,d}}$, which use $T_{\text{0}}+\Delta T_{\text{sc}}$ and $T_{\text{0}}$ as boundary conditions. See text for modeling details.}
\label{trend}
\end{figure}

Raw data of $I_{\text{th}}$ and $\partial^{2}I/\partial V^{2}$ as well as their ratios $R=I_{\text{th}}/\left(\partial^{2}I/\partial V^{2}\right)$ are shown in Figs.~\ref{data}b and \ref{stack}a. In order to extract $\Delta T_{\text{s}}$ and $\Delta T_{\text{d}}$ from the raw data, each $R$ data point is assigned a temperature rise by numerically solving Eq.~(\ref{Signal Ratio}) using Mathematica. Histograms of the numerically determined $\Delta T_{\text{s,d}}$ data (see Fig.~\ref{stack}b) are then used to calculate a mean $\Delta T_{\text{s,d}}$ as well as to define error bars, which encompass two thirds of the data points and are not necessarily symmetric about the mean. The data outside the error bars are typically systematic errors associated with the divergence of $R$ when $\partial^{2}I/\partial V^{2}$ approaches zero. Using the mean value from the histogram to plot Eq.~\ref{Signal Ratio} on top of the original data provides an additional confirmation that the resulting curve agrees with the data. $\Delta T_{\text{s,d}}$ determined from this analysis and plotted versus heating current at a constant background temperature exhibit a consistent trend and a good signal-to-error ratio (see Fig.~\ref{trend}).

To confirm whether the results for $\Delta T_{\text{s}}$ and $\Delta T_{\text{d}}$ in Fig.~\ref{trend} are reasonable, we performed finite element modeling. We used  COMSOL's Multiphysics Electromagnetics and Heat Transfer modes complemented by a custom e-ph coupling term of the form $\Sigma \left(T_{\text{e}}^{n}-T_{\text{ph}}^{n} \right)$, where the coupling constant, $\Sigma$, and the exponent, $n$, depend on the material and the e-ph interaction mechanism. Distinct electron ($T_{\text{e}}$) and phonon ($T_{\text{ph}}$) temperatures are modeled explicitly. In addition to the nanowire itself, the model includes the metallic leads in their entirety, the 100 nm SiO$_{\text{x}}$ layer, the top $10\mu$m of the n-doped Si wafer, and the embedded quantum dot, whose precise position along the nanowire was found via SEM imaging. In the metallic leads and nanowire, we use the electrical conductivity, $\sigma$, and phonon mean free path, $\lambda$, as input parameters and calculate the electronic and phononic thermal conductivities via the Wiedemann-Franz law and Debye model, respectively.  

We first calculate the temperature rise in the heated source contact, $\Delta T_{\text{sc}}$, in the absence of a nanowire. For the source and drain metal leads we use $\sigma_{\text{Au}}=4.05 \times 10^{7}$ $\Omega^{-1}\text{m}^{-1}$ (measured), $\lambda_{\text{Au}}=100$ nm (comparable to the smallest dimension of the Au heating wire), $\Sigma_{\text{Au}}=1.8 \times 10^{9}$ $\text{W}\text{m}^{-3}\text{K}^{-5}$, (averaged value from Refs.~\cite{Henny97} and \cite{Giazotto06}), and $n_{\text{Au}}=5$ (from Refs.~\cite{Henny97} and \cite{Giazotto06}). The dashed line in Fig.~\ref{trend} shows the predicted $\Delta T_{\text{sc}}$ as a function of heating current. The corresponding temperature rise in the drain contact, $\Delta T_{\text{dc}}$, is found to be negligible and is not shown in the figure. After establishing these temperatures in the metallic source and drain contacts, we turn our attention to the nanowire temperatures.

Assuming that the total thermal conductance of the nanowire is dominated by the large electronic thermal resistance of the quantum dot, one would expect that $\Delta T_{\text{s}} \approx \Delta T_{\text{sc}}$ and $\Delta T_{\text{d}} \approx \Delta T_{\text{dc}}\approx 0$. However, this expectation is not consistent with our experimental results, which show that $\Delta T_{\text{s}}$ is only about half as large as the calculated $\Delta T_{\text{sc}}$, and $\Delta T_{\text{d}}$ is, in fact, on the same order as $\Delta T_{\text{s}}$ (see Fig.~\ref{trend}). Therefore, in addition to electronic heat flow, the nanowire must influence $\Delta T_{\text{d}}$ by some other mechanism; we consider Joule heating and e-ph interaction. Additional modeling results show that Joule heating inside the nanowire is too small to explain the observed $\Delta T_{\text{d}}$. Instead, we hypothesize that e-ph interaction inside the nanowire allows the electron gas to couple to phononic heat flow. Conceptually, heated electrons inside the nanowire between the source contact and the quantum dot give thermal energy to the nanowire's phonons via e-ph collisions, thereby decreasing $\Delta T_{\text{s}}$. These heated phonons pass through the quantum dot much easier than electrons and carry heat along the nanowire. On the drain side of the dot, warm phonons couple to cool electrons, thereby increasing $\Delta T_{\text{d}}$.

To perform a qualitative test of our hypothesis, we used the previously calculated electron and phonon temperatures of the metallic source and drain contacts as boundary conditions for the nanowire and calculated $\Delta T_{\text{s,d}}$ in the presence of e-ph coupling inside the nanowire. We varied the parameters $\sigma_{\text{nw}}$, $\lambda_{\text{nw}}$, $\Sigma_{\text{nw}}$, and $n_{\text{nw}}$ (we used integer values only) to reproduce the experimental $\Delta T_{\text{s}}$ and $\Delta T_{\text{d}}$ data in Fig.~\ref{trend}. The parameters $\sigma_{\text{nw}}=8.7 \times 10^{3}$ $\Omega^{-1}\text{m}^{-1}$, $\lambda_{\text{nw}}=100$ nm (comparable to the nanowire diameter), $\Sigma_{\text{nw}}=8.4 \times 10^{10}\text{ W}\text{m}^{-3}\text{K}^{-1}$, and $n_{\text{nw}}=1$ provide a good approximation of the dependence of $\Delta T_{\text{s}}$ and $\Delta T_{\text{d}}$ on $I_{\text{H}}$. Our $\sigma_{\text{nw}}$ is comparable to values found in Ref.~\cite{Hansen05}, and, for a $2\times 10^{17}\text{ cm}^{-3}$ carrier concentration, $\Sigma_{\text{nw}}=2.6\times 10^{6}\text{ eV}\text{s}^{-1}\text{K}^{-1}\text{carrier}^{-1}$, which is within two orders of magnitude of the value reported for a trench-type quantum wire etched in a 2DEG\cite{Bird02}. We conclude that our measured values of $\Delta T_{\text{s}}$ and $\Delta T_{\text{d}}$ are consistent with a model where e-ph coupling allows significant heat flow past the quantum dot into the electron gas on the nanowire's drain side. We caution that $\Sigma_{\text{nw}}$ and $n_{\text{nw}}$ are codependent fit parameters and that other combinations of the two might also fit our data.

In conclusion, we have developed and demonstrated the use of a quantum dot with very narrow transmission resonances for local thermometry of the electrons entering the quantum dot. We have measured temperature differences as large as 0.74 K across a 15 nm quantum dot at 2.94 K. Strictly speaking, the concept of an electron temperature is only meaningful when the electrons establish a thermal distribution. The good agreement of our data with theory (Figs.~\ref{data}b and \ref{stack})) suggests that in our experiment the assumption of Fermi-Dirac distributions is resonable and that any nonequilibrium effects are relatively small. It is important to note that the length scale of electron thermalization puts an upper bound on the spatial precision to which these measured temperatures can be assigned. Our thermometry results agree with a finite element model that includes e-ph coupling as a pathway for heat flow in nanowires. This insight might be important for thermal managment in nanowire-based applications, such as high-speed transistors, LEDs, and thermoelectric devices. Because of the practical challenges involved, few experiments have quantified e-ph interaction in mesoscopic systems\cite{Bird02}. As demonstrated here, quantum-dot thermometry could be put to use testing quantitative predictions of the strength and temperature dependence of e-ph coupling in nanowires.

This research was supported by ONR, ONR Global, the Swedish Research Council
(VR), the Foundation for Strategic Research (SSF), the Knut and Alice
Wallenberg Foundation, and an NSF-IGERT Fellowship.

\end{document}